\documentclass[sigconf, screen]{acmart}

\AtBeginDocument{%
  }

\setcopyright{cc}
\setcctype{by}
\copyrightyear{2026}
\acmYear{2026}
\acmDOI{10.1145/3767308.3835897}

\acmConference[MM '26]{Proceedings of the 34th ACM International Conference on Multimedia}{November 10--14, 2026}{Rio de Janeiro, Brazil}
\acmBooktitle{Proceedings of the 34th ACM International Conference on Multimedia (MM '26), November 10--14, 2026, Rio de Janeiro, Brazil}

\acmISBN{979-8-4007-2213-4/2026/11}

\usepackage{enumitem}
\usepackage{caption}
\usepackage{graphicx}
\usepackage{booktabs}
\usepackage{multirow}
\usepackage{float} 
\usepackage{subcaption}
\usepackage{xcolor}
\usepackage{xspace} 
\usepackage{algorithm}
\usepackage{algorithmic}

\newcommand{\ourmethod}{\textsc{MUSE}\xspace}

\begin{document}

\title[MUSE: A Heterogeneity-Aware Multimedia Search Engine for Mobile SoCs]{\ourmethod: A Heterogeneity-Aware Multimedia Search Engine for Mobile SoCs}

\author{Xinkui Zhao}
\email{zhaoxinkui@zju.edu.cn}
\affiliation{%
  \institution{Zhejiang University}
  \city{Hangzhou}
  \country{China}
}
\affiliation{%
  \institution{Ningbo Global Innovation Center, Zhejiang University}
  \city{Ningbo}
  \country{China}
}
\affiliation{%
  \institution{Zhejiang Key Laboratory of Digital-Intelligence Service Technology}
  \city{Hangzhou}
  \country{China}
}

\author{Qingyu Ma}
\email{maqingyu@zju.edu.cn}
\author{Yifan Zhang}
\correspondingauthor
\email{12451018@zju.edu.cn}
\author{Hengxuan Lou}
\email{12560049@zju.edu.cn}
\author{Sai Liu}
\email{liusai2024@zju.edu.cn}
\affiliation{%
  \institution{Zhejiang University}
  \city{Hangzhou}
  \country{China}
}

\author{Chang Liu}
\email{chang.liu@zju.edu.cn}
\author{Guanjie Cheng}
\email{chengguanjie@zju.edu.cn}
\author{Naibo Wang}
\email{wangnaibo@zju.edu.cn}
\affiliation{%
  \institution{Zhejiang University}
  \city{Hangzhou}
  \country{China}
}
\affiliation{%
  \institution{Ningbo Global Innovation Center, Zhejiang University}
  \city{Ningbo}
  \country{China}
}
\affiliation{%
  \institution{Zhejiang Key Laboratory of Digital-Intelligence Service Technology}
  \city{Hangzhou}
  \country{China}
}

\author{Yueshen Xu}
\email{ysxu@xidian.edu.cn}
\affiliation{%
  \institution{Xidian University}
  \city{Xi'an}
  \country{China}
}

\begin{abstract}
On-device multimedia retrieval is vital for smartphones, enabling applications like cross-modal semantic search and multimodal personal AI agents. However, realizing efficient retrieval on mobile SoCs remains a critical systems challenge. Unlike text-based search, multimedia applications depend on high-dimensional cross-modal embeddings (often exceeding 1024 dimensions) and continuously expanding media streams (e.g., video lifelogs and ambient audio). Mobile devices must process these intensive workloads under strict latency, energy, and memory constraints. Existing vector retrieval systems, designed primarily for static server datasets, are poorly suited for the dynamic execution models of mobile SoCs, where interactive queries must coexist with continuous background media ingestion and index maintenance. To address these challenges, we propose \textbf{\ourmethod}, a \textbf{MU}ltimedia \textbf{S}earch \textbf{E}ngine comprehensively co-designed for mobile SoCs. \ourmethod tackles these issues through an architecture--index co-design. It introduces a three-stage asynchronous execution pipeline that overlaps DMA transfers, NPU matrix computation, and hardware-vector data adaptation to overcome the memory wall of high-dimensional embeddings. Furthermore, it redesigns the Inverted File Index (IVF) and introduces workload-aware heterogeneous scheduling across the CPU, GPU, and NPU for hybrid interactive-and-ingestion workloads. Evaluated on Snapdragon 8-series SoCs using real-world multimodal datasets, \ourmethod{} improves query throughput by up to \(1.4\times\) at matched recall, achieves up to \(7\times\) faster index construction, and delivers up to \(6\times\) higher insertion throughput under concurrent streaming. Crucially, its accelerator-native design ensures strict physical compliance, capping peak device temperatures at \(38^\circ\text{C}\) and reducing total energy consumption by up to \(4.6\times\) compared to CPU-bound baselines.
\end{abstract}

\ccsdesc[500]{Information systems~Information retrieval}
\ccsdesc[500]{Information systems~Multimedia information systems}
\ccsdesc[500]{Computer systems organization~Embedded systems}

\keywords{Multimedia Retrieval, Cross-modal Search, Mobile SoCs, Heterogeneous Computing, Vector Search, Multimodal AI Agents}

\renewcommand{\shortauthors}{Xinkui Zhao et al.}
\maketitle 

\section{Introduction}

Emerging mobile applications, including multimodal personal agents and augmented reality (AR) glasses, increasingly depend on semantic access to large personal multimedia collections, such as photos, ambient audio, and continuous screen recordings~\cite{long2025seeing,shin2025enhancing,mei2024aios,bovo2025embardiment,wang2025karma,tang2025does,zhang2025sortinghat}. To enable cross-modal reasoning (e.g., ``Find the video clip where I saw a red car while listening to jazz''), preserve user privacy, and reduce response latency, these heterogeneous multimedia streams are typically encoded into compact feature vectors by on-device Vision-Language Models (VLMs) and stored in local vector databases for semantic similarity retrieval~\cite{weerasekara2025privacy,zhang2025lsrp}.

\begin{figure}[t]
    \centering
    \includegraphics[width=1\linewidth]{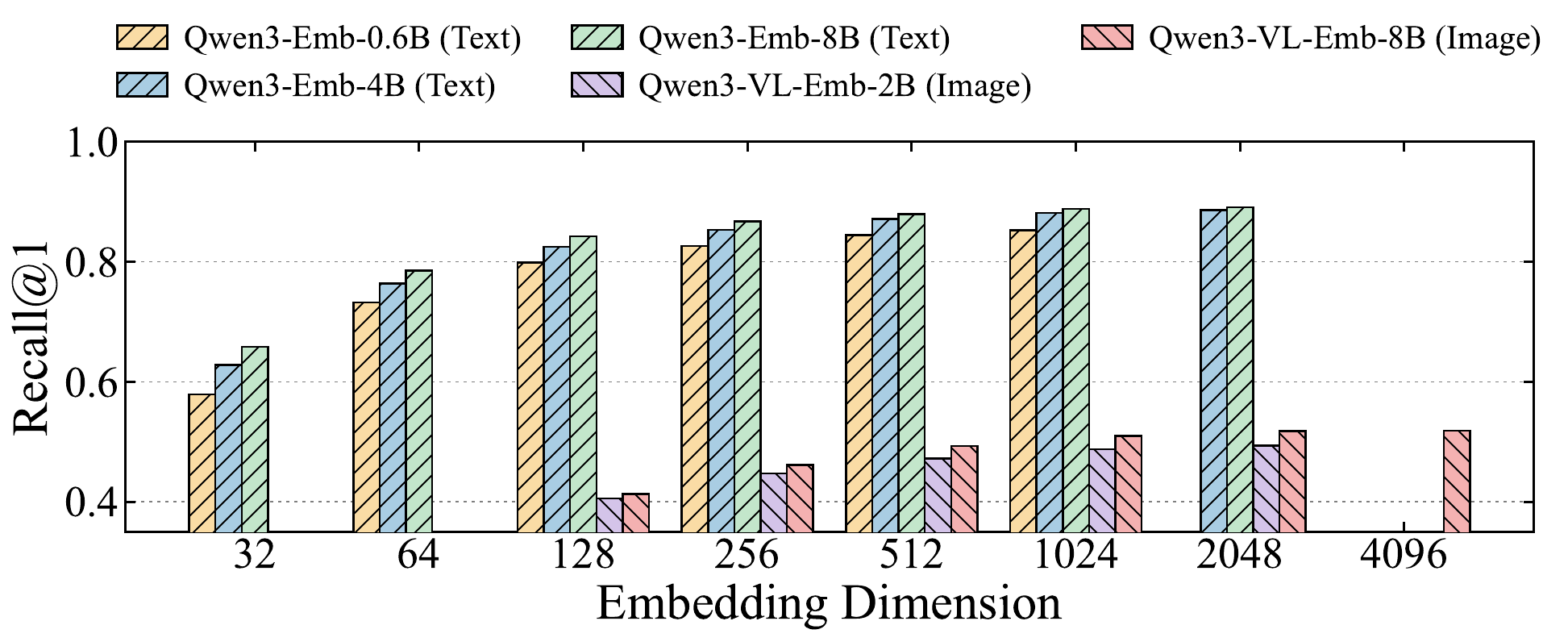}
    \caption{The impact of embedding dimensionality on retrieval performance across encoder types.}
    \label{fig:background}
\end{figure}

However, efficient on-device multimedia retrieval introduces unique systems challenges that far exceed standard text-based search. Modern multimedia workloads rely on exceptionally dense, high-dimensional cross-modal embeddings—often exceeding 1024 or even 4096 dimensions—to retain fine-grained perceptual details and complex semantic alignments across vision, audio, and text spaces~\cite{girdhar2023imagebind, huang2024survey, radford2021learning, li2026qwen3vlembeddingqwen3vlrerankerunifiedframework}. As shown in Figure~\ref{fig:background}, competitive visual and multimodal retrieval accuracy heavily depends on such high-dimensional representations. The challenge is further amplified by the highly dynamic nature of multimedia on mobile devices: personal media libraries continuously expand via uninterrupted background data streams (e.g., continuous frame sampling, audio lifelogging), necessitating continual cross-modal retrieval capabilities~\cite{zhang2023c2mr}. Consequently, on-device multimedia retrieval systems face a severe hybrid workload: they must sustain high-throughput background vector insertions (media ingestion) and ongoing index maintenance, while simultaneously providing low-latency, interactive cross-modal search responses.

These workload demands expose the limitations of existing on-device retrieval engines. While the multimedia community has focused on algorithmic accelerations such as cross-modal hashing~\cite{liu2023multi} and modality-adaptive model pruning~\cite{lyu2025puma}, systems-level execution on mobile systems-on-chip (SoCs) remains a critical bottleneck. Most vector search systems are designed for server-class hardware~\cite{douze2025faiss, wang2021milvus} and struggle with mobile hybrid workloads that interleave high-frequency insertions with low-latency searches, while CPU-based pipelines lack the parallel throughput for real-time responsiveness. Offloading to mobile Neural Processing Units (NPUs) is a natural remedy, yet our empirical study reveals a surprising \textit{performance paradox}: naïvely ported implementations of state-of-the-art vector search algorithms often perform worse than optimized CPU execution---an unoptimized NPU implementation of HNSW (Hierarchical Navigable Small World) incurs up to a $37\times$ latency penalty at high dimensions, and even compute-oriented FLAT (exhaustive search) and hybrid IVF workloads underperform the CPU by $2.1\times$--$3.6\times$ (Section~\ref{sec:moti}). Our profiling attributes this paradox to a deep mismatch with mobile SoC execution models: irregular graph traversals and pointer-chasing map poorly to regular, block-oriented execution of NPUs, while fragmented data movement and repeated layout conversions introduce substantial overhead.

To bridge these gaps, we present \textbf{\ourmethod}, a heterogeneous multimedia search engine tailored for mobile SoCs. \ourmethod combines three complementary techniques: \textbf{(1) an NPU-centric retrieval pipeline} that refactors retrieval computations into dense linear algebra and overlaps DMA transfers, HMX computation, and HVX layout conversion within a zero-copy shared-memory design; \textbf{(2) a hardware-aligned IVF design} that restructures index parameters to match NPU-friendly execution granularity, allowing centroid routing, candidate scoring, insertion, and maintenance to map cleanly onto accelerator kernels; and \textbf{(3) workload-aware heterogeneous scheduling} that dynamically maps interactive queries, batch retrieval, and background media ingestion across the CPU, GPU, and NPU using profiling-guided policies.

In summary, this paper makes the following contributions:
\begin{itemize}[leftmargin=*]
    \item To the best of our knowledge, \ourmethod is the first on-device multimedia retrieval system that systematically co-designs high-dimensional vector search for heterogeneous smartphone processors, including the CPU, GPU, and NPU.
    \item We design an NPU-centric retrieval pipeline that refactors vector retrieval into dense computations and overlaps asynchronous DMA transfers, HMX matrix execution, and HVX layout conversion, reducing data adaptation and cross-processor overheads.
    \item We develop a hardware-aligned IVF design and a workload-aware heterogeneous scheduler for multimedia workloads, enabling efficient query processing, vector insertion, and index maintenance across the CPU, GPU, and NPU.
    \item Extensive experiments on commercial flagship mobile SoCs show that \ourmethod improves query throughput by up to \(1.4\times\) at matched recall, accelerates index construction by up to \(7\times\), and increases insertion throughput by up to \(6\times\) under hybrid query-update workloads, while capping peak device temperature at \(38^\circ\text{C}\) and reducing total energy consumption by up to \(4.6\times\).
\end{itemize}

\section{Background}

\begin{figure}[t]
    \centering
    \includegraphics[width=1\linewidth]{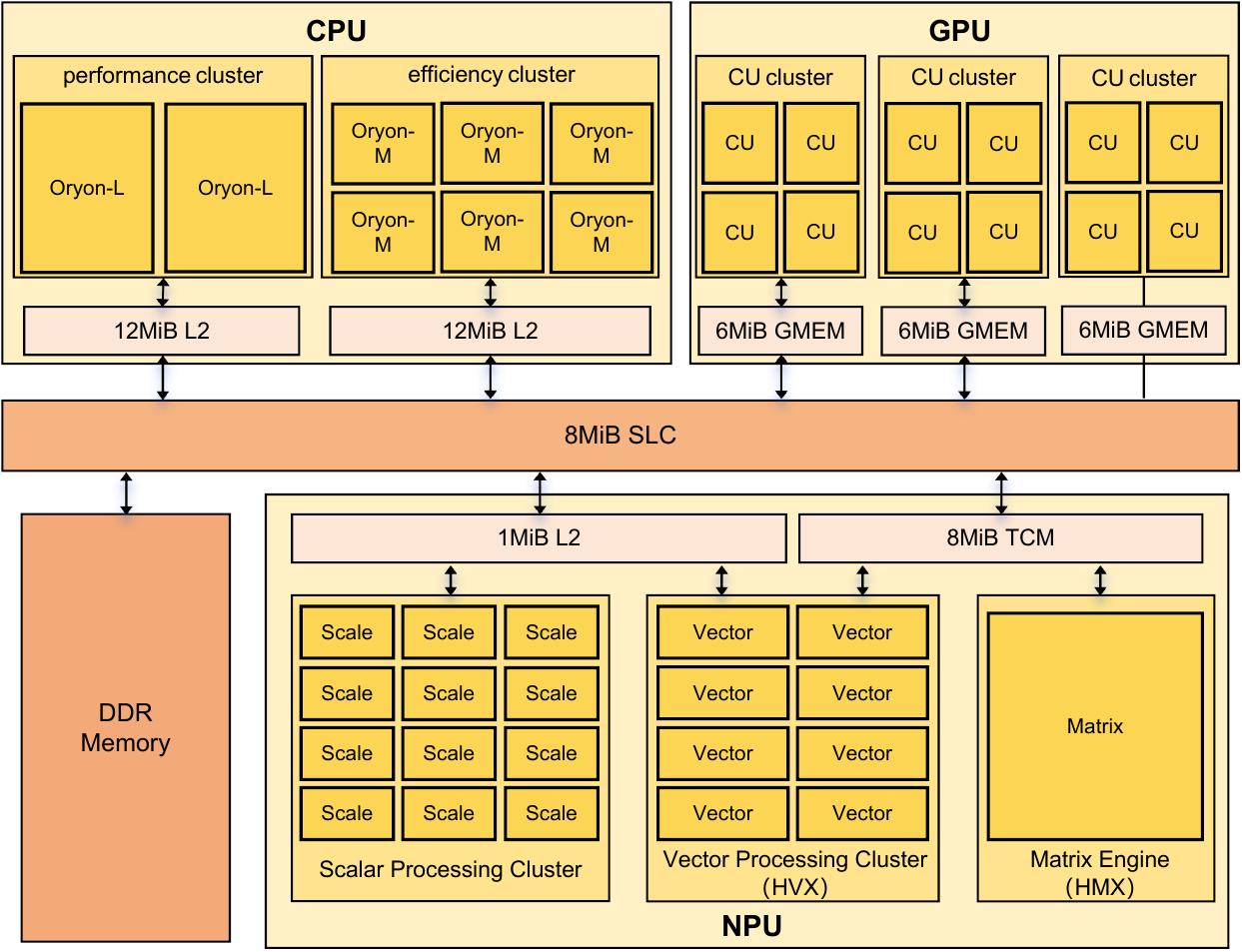}
    \caption{The architectural blueprint of the Snapdragon 8 Elite Gen 5, highlighting heterogeneous compute units (CUs) and the shared memory hierarchy. Oryon-L denotes Qualcomm's custom high-performance CPU core; the Hexagon NPU integrates the HVX vector unit and HMX matrix engine, backed by a tightly coupled memory (TCM).}
    \label{fig:soc}
\end{figure}

Modern mobile systems-on-chip (SoCs) integrate heterogeneous processors—CPUs, GPUs, and NPUs—that expose markedly different execution models and preferred data layouts, making cross-processor collaboration a central systems challenge~\cite{zheng2025review}. Unlike server-class discrete architectures, mobile SoCs employ a unified or tightly shared memory hierarchy, which reduces explicit data-copy overhead and enables low-latency data sharing across processors. Meanwhile, mobile accelerators rely on small local buffers or tightly coupled memories to stage high-bandwidth data, so performance depends critically on how data movement, layout transformation, and computation are orchestrated.

This general design is exemplified by the Snapdragon SoC shown in Figure~\ref{fig:soc}, whose NPU subsystem is based on Qualcomm's Hexagon architecture~\cite{qualcomm_hexagon_arch}. Hexagon integrates scalar control logic, a vector unit (HVX), and a matrix engine (HMX), together with a small cache hierarchy and an 8~MiB tightly coupled memory (TCM) for staging tensor tiles. In this architecture, HVX supports vector-processing tasks such as data conversion, layout packing, and lightweight tensor manipulation, whereas HMX provides high-throughput matrix computation for similarity scoring.

\begin{figure}[t]
    \centering
    \includegraphics[width=1\linewidth]{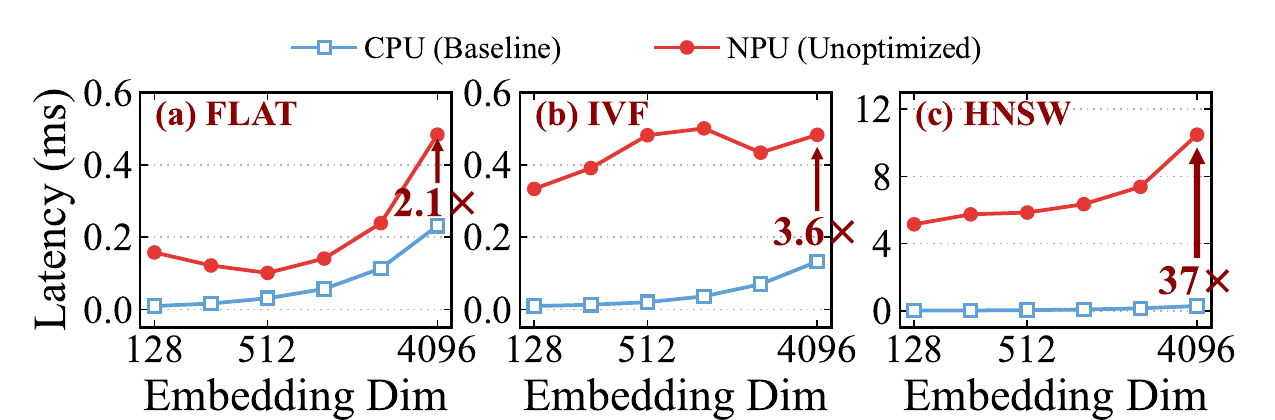}
    \caption{Comparison of retrieval latency between mobile CPU and an unoptimized NPU across different indexing algorithms.}
    \label{fig:motivation}
    
    \includegraphics[width=0.85\linewidth]{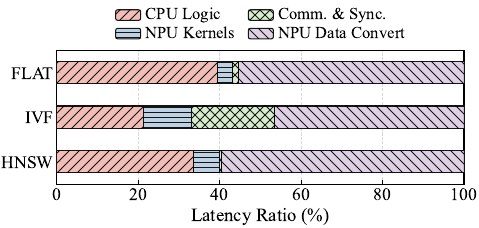}
    \caption{Latency breakdown of unoptimized multimedia vector search on a mobile NPU.}
    \label{fig:analysis}
\end{figure}

\section{Motivation and Analysis}
\label{sec:moti}

\noindent \textbf{Experiment: Impact of Mobile SoCs on Multimedia Vector Search.}
Server-oriented vector search algorithms such as HNSW~\cite{malkov2018efficient} and IVF~\cite{douze2025faiss} assume deep cache hierarchies and abundant memory bandwidth—assumptions that break down under the tight resource constraints of mobile SoCs. To quantify the consequences, we conduct a microbenchmark of representative methods—FLAT, IVF, and HNSW—on a commodity mobile NPU. A major technical hurdle is that commercial mobile NPUs are predominantly exposed via black-box model-level APIs, precluding fine-grained operator control~\cite{hao2025scaling}. To circumvent this, we bypass the standard inference engine and utilize undocumented low-level SDK interfaces to construct custom matrix execution kernels. We define this direct, synchronous implementation as our \textit{Unoptimized NPU} baseline. All benchmarks utilize 1024-dimensional embeddings representative of modern vision–language models.

As shown in Figure~\ref{fig:motivation}, retrieval latency on this naïve NPU implementation paradoxically increases by up to $37\times$ relative to optimized mobile CPU execution. To identify the source of this degradation, Figure~\ref{fig:analysis} decomposes per-query latency into four primary components: host-side control flow and index traversal (CPU Logic), data type conversion and layout transformation on the NPU (NPU Data Convert), inter-processor communication and synchronization overhead (Comm.~\&~Sync.), and the hardware-accelerated distance computation itself (NPU Kernels).

\noindent \textbf{Analysis.} Our microbenchmarks indicate that the poor performance of existing vector search methods on mobile NPUs stems primarily from three architectural mismatches with the target on-device multimedia retrieval workload.

\textit{(1) Mismatch between irregular search algorithms and accelerator execution models.}
A substantial portion of the latency overhead stems from the mismatch between graph-based search algorithms and NPU-oriented execution models. As shown in Figure~\ref{fig:analysis}, methods such as HNSW depend on irregular graph traversal and pointer chasing, which are well suited to CPUs with branch prediction and cache support but map poorly to mobile NPUs, which are typically optimized for regular, block-structured tensor computation. Consequently, irregular accesses to non-contiguous embeddings result in frequent stalls and increased host-side coordination overhead.

\textit{(2) Costly data conversion and layout adaptation.}
Another major source of overhead is the mismatch between the embedding representations used by the retrieval stack and the numeric formats preferred by mobile accelerators. Although embeddings are commonly stored and processed in FP32 on the CPU side, NPUs typically favor lower-precision formats such as FP16 or INT8. As a result, data conversion and layout adaptation can account for a substantial fraction of end-to-end latency, diminishing the benefits of accelerator offloading when performed naïvely.

\textit{(3) Limited support for fine-grained memory orchestration.}
Finally, current mobile accelerator stacks provide limited support for the fine-grained and sparse access patterns induced by IVF and graph-based indices. As a result, the CPU must frequently participate in staging intermediate data, managing additional software-controlled transfers, and synchronizing between shared storage and accelerator-local memory. These communication overheads become increasingly significant as embedding dimensionality and workload intensity grow.

\section{\ourmethod Design}

\subsection{Overview}
We introduce \ourmethod, a collaborative heterogeneous multimedia search engine custom-built for mobile SoCs. \ourmethod{} combines three complementary techniques: (1) an NPU-centric computation pipeline, (2) a hardware-aligned IVF index design, and (3) a workload-aware heterogeneous scheduler.

\begin{figure*}[t]
  \centering
  \includegraphics[width=\linewidth]{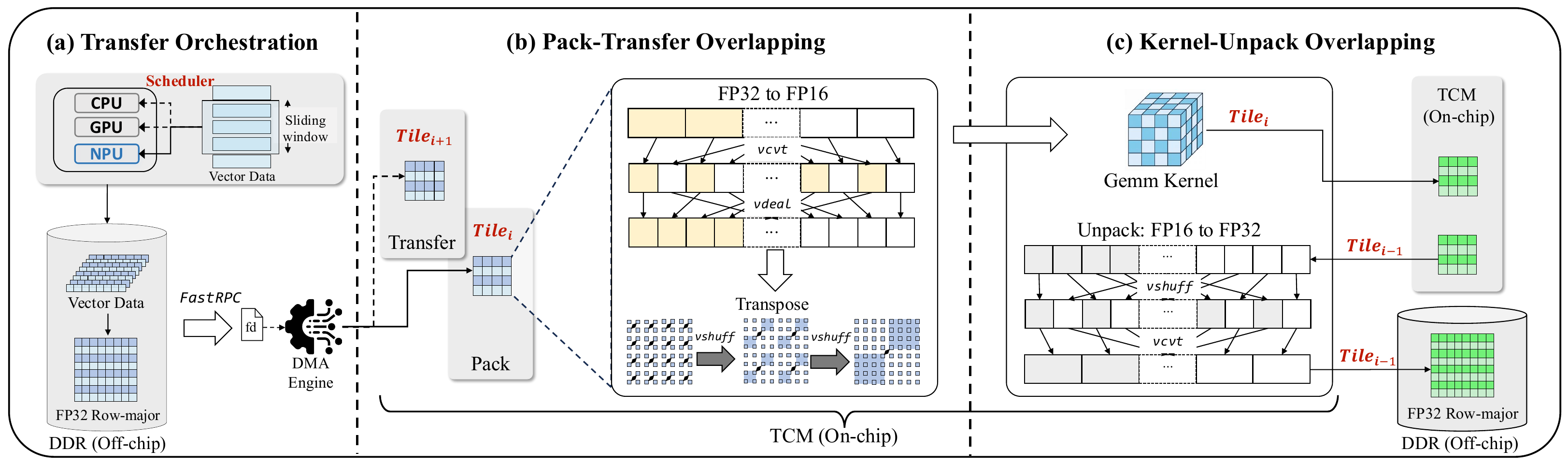}
  \caption{End-to-end heterogeneous scheduling and asynchronous NPU execution pipeline in \ourmethod. (a) Transfer orchestration using a sliding-window scheduler and zero-copy memory access. (b) Pack-transfer overlapping, hiding on-the-fly FP32$\rightarrow$FP16 layout conversion behind DMA fetches. (c) Kernel-unpack overlapping, running GEMM computation concurrently with FP16$\rightarrow$FP32 result reconstruction.}
  \label{fig:conv}
\end{figure*}

\subsection{NPU-centric Computation Pipeline}

Although modern mobile SoCs expose heterogeneous processors — including CPUs, GPUs, and NPUs—efficient high-dimensional multimedia retrieval fundamentally depends on optimizing the NPU execution path for throughput-oriented scoring. As analyzed in Section~\ref{sec:moti}, naïve offloading typically serializes DMA transfers, layout conversions, and matrix operations, leaving the NPU's computation engine idle for extended periods. To sustain hardware saturation, \ourmethod implements an asynchronous execution pipeline (Figure~\ref{fig:conv}) that structurally decouples memory I/O, format adaptation, and matrix arithmetic across three concurrent phases.

\paragraph{(a) Transfer Orchestration.}
A primary source of latency in mobile heterogeneous systems is the boundary crossing between the host CPU and the accelerator. \ourmethod addresses this cross-processor communication overhead on two fronts. First, to amortize the high, fixed runtime penalty of \texttt{FastRPC} context switches, the system batches multiple micro-tile retrieval operations into a single collective accelerator invocation. 

Second, standard \texttt{FastRPC} and NPU execution frameworks typically force explicit host-to-host memory allocation and buffer duplication to prepare accelerator-specific layouts before invocation. As shown in Figure~\ref{fig:conv}(a), \ourmethod bypasses this host-side overhead by maintaining the global vector database in a host-friendly FP32 row-major layout within off-chip DDR. Instead of copying tensor payloads into dedicated NPU RPC buffers, the host directly shares memory mappings using standard mobile \texttt{dma-buf} and ION (Android's memory allocation framework) file descriptors (\texttt{fd})~\cite{kang2025puzzle, van2016drammer}. This zero-copy design completely eliminates OS-level memory duplication; while the NPU's DMA engine inherently must pull the data into its TCM for physical execution, the data-movement burden is shifted entirely to the hardware DMA, saving crucial CPU cycles and memory bandwidth.

\paragraph{(b) Pack-Transfer Overlapping.}
Retrieval scoring typically follows an $AB^T$ GEMM computation pattern between a query batch and candidate embeddings, meaning the database matrix ($B$) must be transposed. Because the NPU's HMX strictly consumes FP16 tile-major inputs, the host's FP32 row-major embeddings must be converted. Performing this conversion on the CPU incurs substantial additional memory traffic and requires temporary intermediate buffers, which are prohibitive under the tight memory bandwidth budgets of mobile devices. 

Instead, \ourmethod performs this adaptation in-place on the HVX. Even after IVF pruning, multimedia candidate pools far exceed the limited capacity of the on-chip TCM, requiring data to be streamed in tiles. As detailed in Figure~\ref{fig:conv}(b), the pipeline hides this streaming overhead by overlapping memory I/O with data formatting. While the DMA engine transfers the next data block ($Tile_{i+1}$) into TCM, an asynchronous HVX thread repacks the previously fetched data. The process maps directly to vector ISA semantics: the thread downcasts the precision from FP32 to FP16 using \texttt{vcvt}, interleaves the sub-blocks using \texttt{vdeal}, and applies \texttt{vshuff} to locally transpose the elements into a tile-major layout.

\paragraph{(c) Kernel-Unpack Overlapping.}
To sustain peak matrix throughput, the computation must not block while waiting for output formatting. \ourmethod achieves this through the final pipeline stage (Figure~\ref{fig:conv}(c)), employing a background worker pool to overlap execution.

During a steady-state cycle, the HMX unit executes the dense GEMM kernel on the current tile ($Tile_i$) residing in TCM. Simultaneously, a separate HVX worker thread processes the output of the previous tile ($Tile_{i-1}$). It reverses the layout using \texttt{vshuff}, upcasts the results back to FP32 using \texttt{vcvt}, and writes the final row-major scores back to off-chip DDR~\cite{qualcomm_hexagon_hvx}. By maintaining independent partitioned ping-pong buffers in the TCM and synchronizing thread progress via lightweight worker tokens, \ourmethod pushes both memory latency and data adaptation overhead completely out of the critical path, allowing the matrix engine to achieve near-peak utilization during large-batch multimedia retrieval.

\subsection{Index Design}

\begin{table}[t]
\centering
\caption{Representative vector indexes and their limitations for high-dimensional embedding retrieval on mobile SoCs.}
\renewcommand{\arraystretch}{1.1}
\resizebox{\columnwidth}{!}{%
\begin{tabular}{c|c|c}
\toprule
\textbf{Algorithm} & \textbf{Optimized for} & \textbf{Limitation on SoC} \\
\midrule
FLAT~\cite{weber1998quantitative}                 & Exact search          & High compute and memory-bandwidth cost \\
HNSW~\cite{malkov2018efficient}                 & Cache-rich CPUs       & Irregular graph traversal and pointer chasing \\
IVF~\cite{jegou2010product}                  & CPU / GPU             & DRAM footprint and scattered candidate probes \\
Faiss IVF--PQ (CPU)~\cite{johnson2019billion}  & CPU                   & Host-side compute overhead, DRAM-bound access \\
Faiss IVF--PQ (GPU)~\cite{johnson2019billion}  & Nvidia GPU            & CUDA-specific implementation, high power / thermal cost \\
\bottomrule
\end{tabular}%
}
\label{tab:vdb_comparison}
\end{table}

\ourmethod{} adopts IVF as its retrieval backbone rather than graph-based designs such as HNSW (Table~\ref{tab:vdb_comparison}). FLAT search is excluded because its compute and memory-bandwidth costs become prohibitive at realistic database scales. Compared with HNSW, IVF is better suited to mobile NPUs: HNSW relies on graph traversal, pointer chasing, and irregular control flow~\cite{malkov2018efficient,ootomo2024cagra}, whereas IVF decomposes retrieval into centroid routing and candidate scoring, both of which can be expressed as dense batched vector operations. IVF is also more amenable to dynamic multimedia workloads, as insertions require only centroid assignment and list updates rather than graph traversal and neighbor maintenance.

This structure further enables hardware-aware mapping to the NPU. On Hexagon, both centroid routing and candidate scoring can be organized as GEMM-style kernels executed by HMX. The HMX matrix engine achieves its highest efficiency with a tile shape of \(32 \times 64 \times 64\) (\(M \times N \times K\)). In the context of IVF retrieval, \(M\) corresponds to the query batch size, \(N\) to the number of centroids or candidate vectors, and \(K\) to the embedding dimensionality. Therefore, we configure the number of IVF clusters (\(N\)) as a multiple of 64, while modern embedding dimensions (e.g., 512 or 1024) already satisfy the alignment requirement along the \(K\) dimension. The query batch dimension (\(M\)) is padded to the nearest multiple of 32. Together, these choices allow the core IVF computations to map cleanly onto HMX tiles with minimal fragmentation.

\subsection{Heterogeneous Scheduler Design}
\label{sec:coord}

\paragraph{Workload-Aware Routing for Media Streams.}
Multimedia retrieval generates a highly diverse spectrum of workloads. As our hardware profiling (Section~\ref{sec:profiling}) shows, mobile processors have distinct efficiency regimes: the NPU provides massive throughput for large matrix operations but incurs fixed \texttt{FastRPC} initialization costs; the CPU handles control flow efficiently but lacks dense compute throughput; the GPU falls in between.

Rather than forcing all requests to the NPU, \ourmethod implements a workload-aware scheduler (Figure~\ref{fig:conv}(a)) to dynamically route tasks based on the specific multimedia use-case:
\begin{itemize}[leftmargin=*]
    \item \textbf{Cross-modal Interactive Query (CPU):} For single-item text-to-image or visual question answering queries, matrix dimensions are too small to amortize NPU invocation overhead. \ourmethod routes these latency-sensitive foreground tasks to the CPU, ensuring instantaneous user feedback.
    \item \textbf{Continuous Media Ingestion (CPU--GPU):} For uninterrupted background media ingestion (e.g., streaming embeddings from screen recordings or audio lifelogging), \ourmethod relies on CPU--GPU co-processing. The GPU executes batched insertions, leaving the NPU entirely free for the agent's primary Vision-Language Model inference~\cite{chu2024mobilevlmv2fasterstronger, liu2024mobilellm}.
    \item \textbf{Periodic Index Rebuild (Full SoC):} Periodic index maintenance over massive personal photo/video libraries generates massive dense matrix operations. Here, \ourmethod engages the CPU, GPU, and NPU concurrently, saturating the SoC's full bandwidth to minimize rebuild time and background thermal strain.
\end{itemize}
Across all modes, branch-heavy operations such as global Top-$K$ reduction are universally retained on the CPU, preventing control-flow divergence on parallel accelerators.

\paragraph{Memory-Efficient Task Orchestration.}
Processing massive personal databases creates a risk of memory exhaustion if a rebuild workload is materialized in memory all at once. \ourmethod prevents this through a windowed batch submission model. 

As depicted in Figure~\ref{fig:conv}(a), the system does not statically partition the dataset across backends. Instead, \ourmethod implements a dynamic pull mechanism governed by a configurable device mask. It maintains a sliding window of fixed-size vector data chunks; worker threads pinned to the enabled CPU, GPU, and NPU autonomously pull tasks from this window upon completing their current workloads. This decentralized design naturally eliminates execution bubbles—since faster processors steal work more frequently—while strictly bounding the peak memory footprint, allowing \ourmethod to safely process arbitrarily large multimedia indexes within tight mobile RAM constraints.

\section{Experimental Results}
\subsection{System Implementation}
We implemented \ourmethod atop a customized mobile AI runtime integrating OpenCL~\cite{munshi2011opencl}, OpenBLAS~\cite{nugteren2018clblast}, the Qualcomm Hexagon DSP SDK, and heavily optimized GGML-based~\cite{ggml2024} tensor kernels. The framework is highly portable across modern heterogeneous architectures and comprises approximately 19K lines of C/C++ code.

\subsection{Experiment Setup}

\begin{figure*}[t]
  \centering
  \includegraphics[width=1\linewidth]{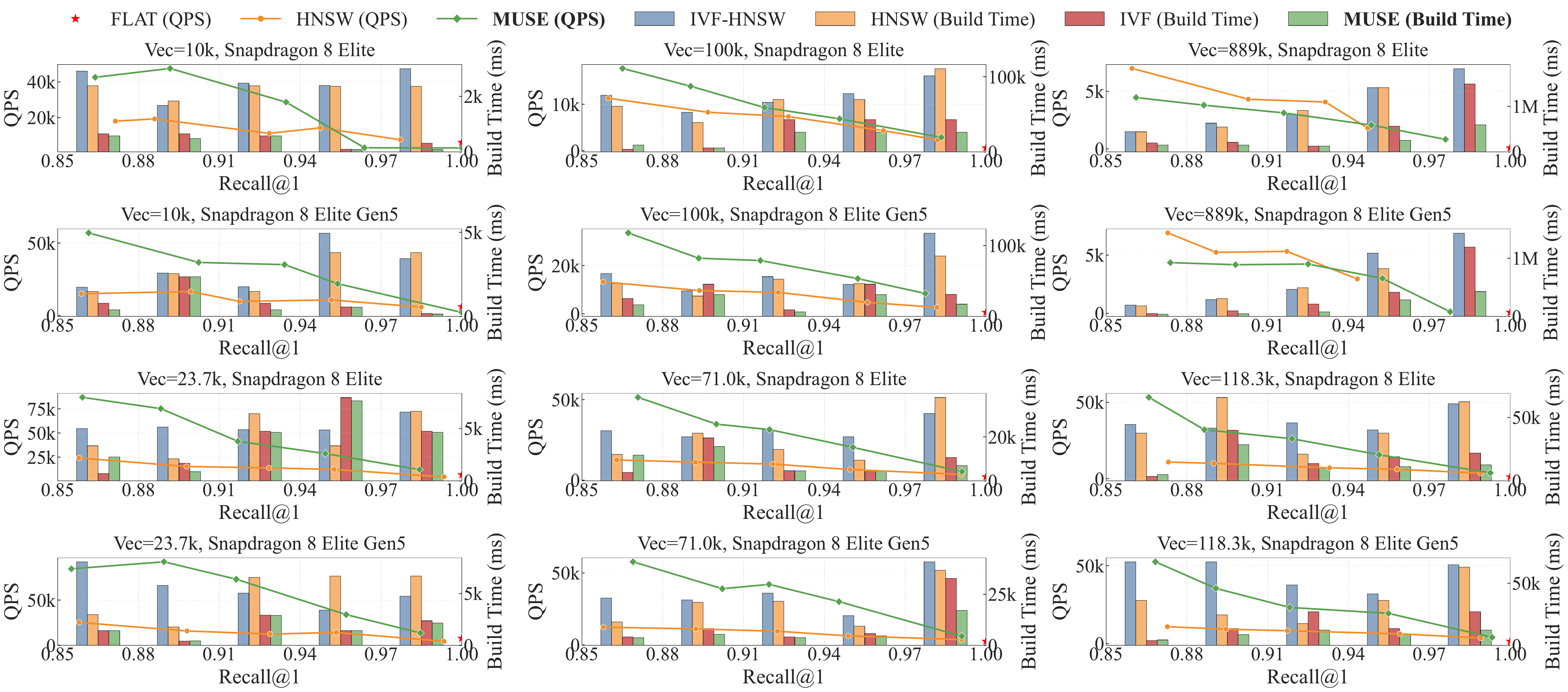}
  \caption{Query performance and index construction/rebuild latency for private on-device search: Recall@1 vs. QPS (left) and build time (right) across diverse multimedia database scales and hardware generations. \ourmethod consistently dominates the Pareto frontier.}
  \label{fig:mainresult}
\end{figure*}

\begin{figure*}[t]
  \centering
  \includegraphics[width=1\linewidth]{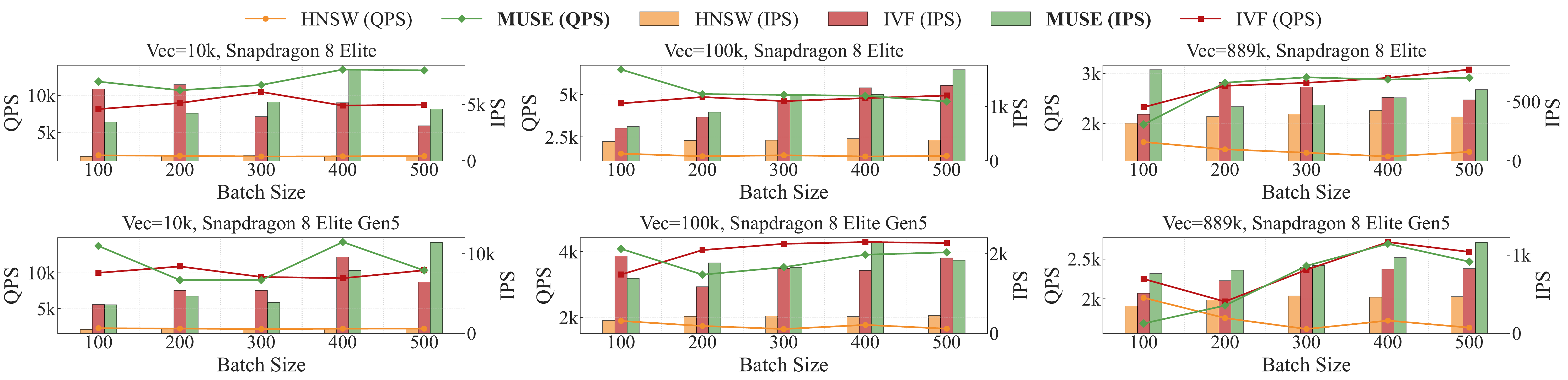}
  \caption{System performance under continuous media ingestion (lifelogging): sustained query throughput (left) and insertion throughput (right) under concurrent querying and ingestion, normalized against batch sizing.}
  \label{fig:hybrid}
\end{figure*}

\textbf{Platform \& Environment:}
We deployed \ourmethod on commercial heterogeneous platforms representing the state-of-the-art in mobile compute: the Qualcomm Cloud Phone platform (Snapdragon 8 Elite) and the Redmi K90 Pro Max (Snapdragon 8 Elite Gen 5). 

\textbf{Baselines:}
We benchmark \ourmethod against four premier vector-retrieval frameworks optimized for mobile endpoints: FLAT (exhaustive search), IVF, HNSW~\cite{malkov2018efficient}, and IVF-HNSW~\cite{baranchuk2018revisiting}. These canonical baselines isolate the core system question of how to efficiently support large-scale multimedia embedding retrieval under mobile latency, energy, and memory constraints; many optimized IVF or graph variants would conflate algorithmic gains with the system effects we aim to measure.

\textbf{Metrics \& Workloads:} Evaluations are conducted using standard high-dimensional embedding metrics: Latency (ms), QPS, IPS (Insertions Per Second), and Recall@K (\%). To ensure practical relevance to the multimedia community, we evaluate using massive cross-modal embeddings extracted via desktop GPUs (RTX 3090Ti):
\begin{itemize}[leftmargin=*, topsep=2pt, itemsep=0pt]
    \item \textbf{Text-to-Text Workload (Baseline):} Derived from HotpotQA~\cite{yang2018hotpotqa} using \texttt{Qwen3-Embedding-4B} ($d=1024$), max scale 889K vectors.
    \item \textbf{Image/Vision Workload (Multimedia):} Derived from MS-COCO~\cite{lin2014microsoft} using the multimodal \texttt{Qwen3-VL-Embedding-8B} ($d=1024$), capturing complex visual semantics, max scale 118.3K vectors.
\end{itemize}
These datasets accurately reflect the continuous, high-dimensional feature spaces of modern Large Multimodal Models (LMMs) deployed in personal lifelogging agents.

\subsection{Query and Index Performance}

We first evaluate the foundational retrieval efficiency and index construction velocity of \ourmethod (Figure~\ref{fig:mainresult}). On small to medium-scale multimedia databases, \ourmethod defines the Pareto frontier across the entire Recall-QPS spectrum, delivering up to a $1.4\times$ speedup at equivalent recall targets compared to state-of-the-art baselines. On large-scale corpora, graph-based HNSW initially shows competence at low recall, but \ourmethod definitively overtakes it as recall demands increase. Crucially, at the highest recall constraints, HNSW fails due to mobile memory exhaustion: high-recall HNSW requires large graph-neighbor storage, candidate/frontier state, and insertion/update metadata, which are less amenable to mobile memory constraints than IVF's contiguous vector storage, whereas \ourmethod remains robust.

In terms of index construction—a vital metric for processing massive personal media archives—\ourmethod demonstrates profound superiority. It achieves up to $7\times$ faster build times compared to HNSW at identical recall targets.

\subsection{Hybrid Search-Update Dynamics}
To emulate a continuously evolving multimodal personal agent, we evaluated system behavior under concurrent querying and media insertion (Figure~\ref{fig:hybrid}). Regardless of the insertion batch size, \ourmethod sustains superior operational QPS, completely mitigating the severe blocking observed in HNSW. Under rigorous concurrent stress tests, \ourmethod achieves a remarkable $6\times$ higher insertion throughput than HNSW, proving its capability to serve as a reliable backbone for real-time multimedia ingestion.

\subsection{Ablation Analysis of the NPU Subsystem}

\begin{figure}[t]
    \centering
    \includegraphics[width=1\linewidth]{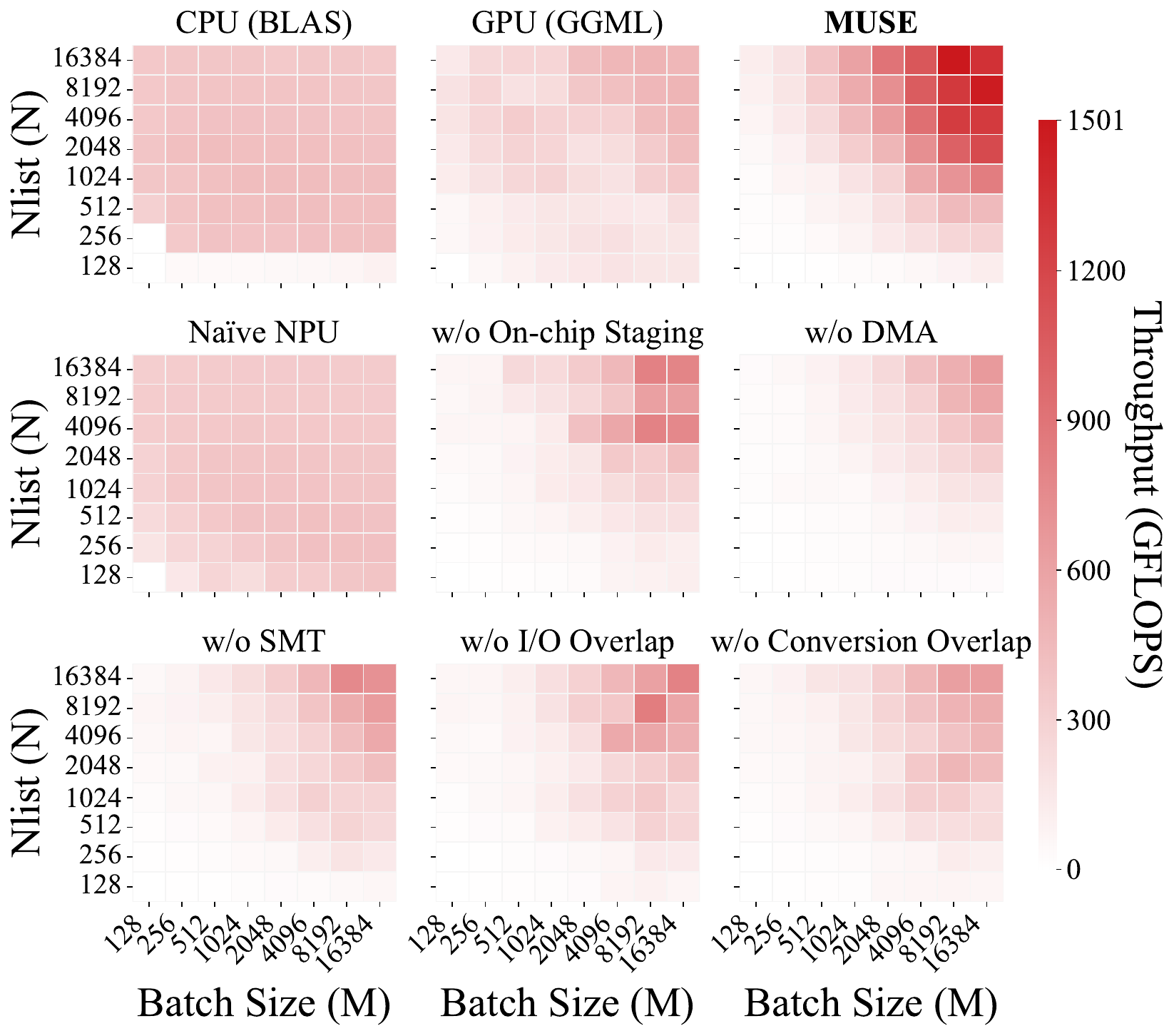}
    \caption{Characterization of GEMM throughput across heterogeneous processors and NPU pipeline ablations.}
    \label{fig:comprehensive_ablation}
\end{figure}

To understand the performance gains of \ourmethod's NPU backend, Figure~\ref{fig:comprehensive_ablation} (Middle \& Bottom rows) breaks down the throughput impact of each design component. The unoptimized \textit{Naïve NPU} baseline underperforms significantly due to stalled execution during memory staging and layout conversions.

Removing the on-chip scratchpad (\textit{w/o On-chip Staging}) forces the NPU to access DDR directly, triggering severe latency penalties. However, merely staging data on-chip is insufficient: relying on synchronous host-managed transfers (\textit{w/o DMA}) yields negligible improvements because the data movement overhead eclipses the computation time. 

The most substantial gains come from \ourmethod's software pipelining. Disabling multithreading (\textit{w/o SMT}) or double-buffered DMA prefetching (\textit{w/o I/O Overlap}) exposes the NPU matrix engine to DDR access latencies. Furthermore, omitting asynchronous HVX data adaptation (\textit{w/o Conversion Overlap}) forces the pipeline to serialize FP16-to-FP32 layout conversions. By maintaining a concurrent three-stage pipeline (DMA I/O, HMX execution, and HVX write-back), \ourmethod completely overlaps memory and format adaptation with computation, achieving near-peak hardware saturation for large matrix sizes.

\subsection{Hardware Profiling and Workload Characterization}
\label{sec:profiling}

To justify our workload-aware scheduling policy (Section~\ref{sec:coord}), we characterize the dense matrix multiplication (GEMM) performance of the CPU, GPU, and NPU across varying dimensions. Since IVF routing and candidate scoring essentially reduce to batched GEMMs ($M \times K$ multiplied by $K \times N$), we fix the embedding dimension ($K$) at 1024—representative of modern multimodal feature spaces—and sweep both the build batch size ($M$) and the cluster count ($N$) from 128 to 16384.

Figure~\ref{fig:comprehensive_ablation} (Top row) reveals distinct hardware efficiency regimes. For small operations (e.g., single-query interactive routing), the CPU delivers the highest throughput, as its execution avoids the memory-mapping and \texttt{FastRPC} kernel-launch overheads that penalize accelerator offloading. However, as matrix sizes expand, CPU performance quickly saturates; its limited vector arithmetic units (e.g., NEON) are structurally ill-equipped to handle the massive dense matrix operations demanded by multimedia workloads. By contrast, the \ourmethod-optimized NPU exhibits a clear scaling behavior. While its throughput is constrained at small dimensions due to the fixed initialization costs of the deep SMT-DMA pipeline, it achieves maximum hardware saturation at large dimensions ($M, N > 1024$).

This performance topology is not a coincidence, but a deliberate system trade-off tailored for \textit{multimedia} retrieval. Text-based search often involves lightweight scalar queries, but multimedia workloads—such as batch-ingesting high-framerate video clips, processing ambient audio streams, and scanning expansive multimodal candidate pools—inherently generate massive $M \times N$ matrix operations. The deep SMT-DMA pipeline and batched \texttt{FastRPC} invocations of \ourmethod incur a fixed initialization penalty that is unfavorable for tiny operations, but they effectively overcome the memory bandwidth limitations for large-scale operations, unlocking sustained high throughput for multimedia scales.

Furthermore, this empirical hardware behavior perfectly justifies \ourmethod's heterogeneous scheduling strategy (Section~\ref{sec:coord}). A naïve ``accelerator-first'' approach that rigidly offloads all tasks to the NPU would penalize latency-sensitive interactive queries due to the aforementioned pipeline startup costs. Conversely, relying solely on the CPU/GPU leaves the SoC's most potent matrix engine idle during heavy background index rebuilding. By exploiting this hardware topology, \ourmethod intelligently routes small-scale interactive tasks to the CPU/GPU, while reserving the hyper-optimized NPU pipeline for the massive index construction and maintenance tasks it was fundamentally designed to dominate.

\subsection{Validation of Hardware-Aware Indexing}
\begin{figure}[t]
    \centering
    \includegraphics[width=0.9\linewidth]{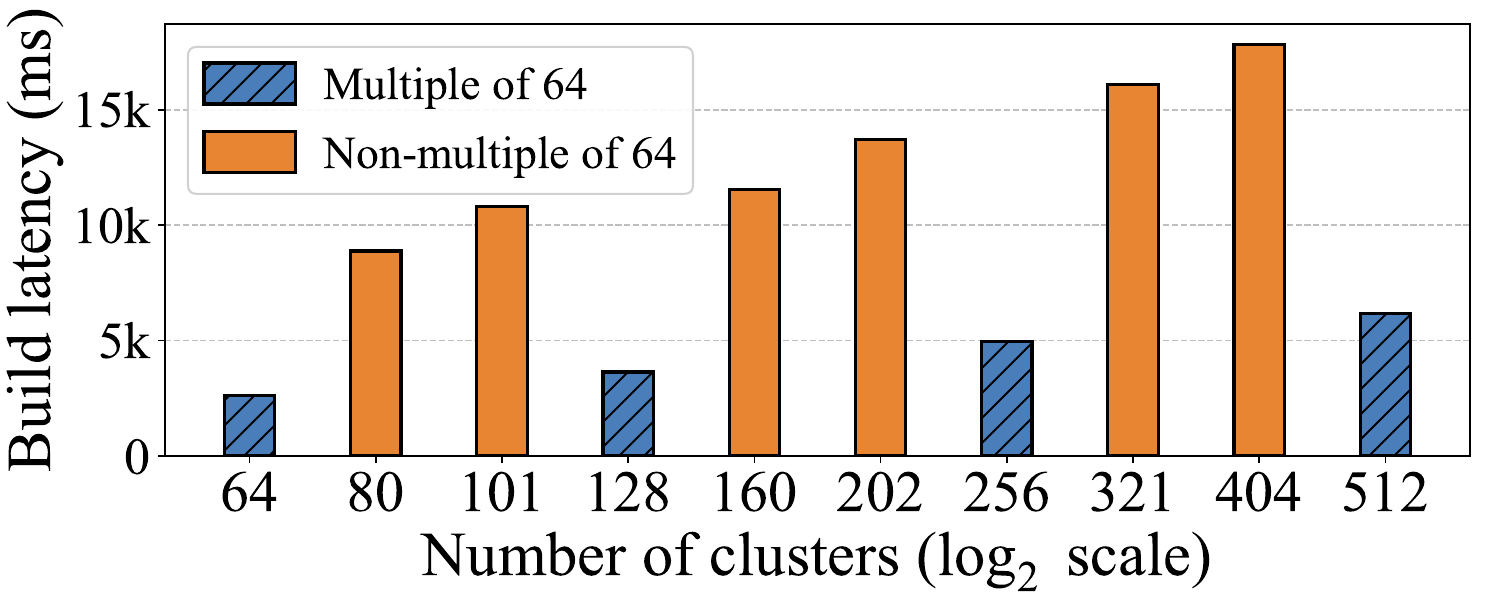}
    \caption{Impact of hardware-aligned IVF cluster configurations on index construction latency.}
    \label{fig:ivf}
\end{figure}

To validate the necessity of aligning IVF clustering dimensions with NPU matrix geometries, we measure index construction latency across a sweep of cluster counts (Figure~\ref{fig:ivf}). The data demonstrates a clear architectural constraint: when cluster counts diverge from multiples of 64, the resulting GEMM operations fragment across partially filled HMX tiles. This misalignment introduces padding overhead and execution inefficiencies, prolonging latencies. Conversely, strictly aligning clusters to $64\times$ intervals yields sharp local minima in construction time, confirming the performance benefits of hardware-software structural co-design.

\subsection{Scalability with High Dimensionality}
As multimodal foundation models evolve, embedding dimensionality increases to capture richer semantics (e.g., $d=4096$ in modern LMMs). High-dimensional retrieval exacerbates memory bandwidth bottlenecks and cache thrashing on mobile CPUs. To evaluate \ourmethod's scalability, we measure end-to-end execution time across embedding dimensions up to $d=4096$.

\begin{figure}[t]
    \centering
    \includegraphics[width=1\linewidth]{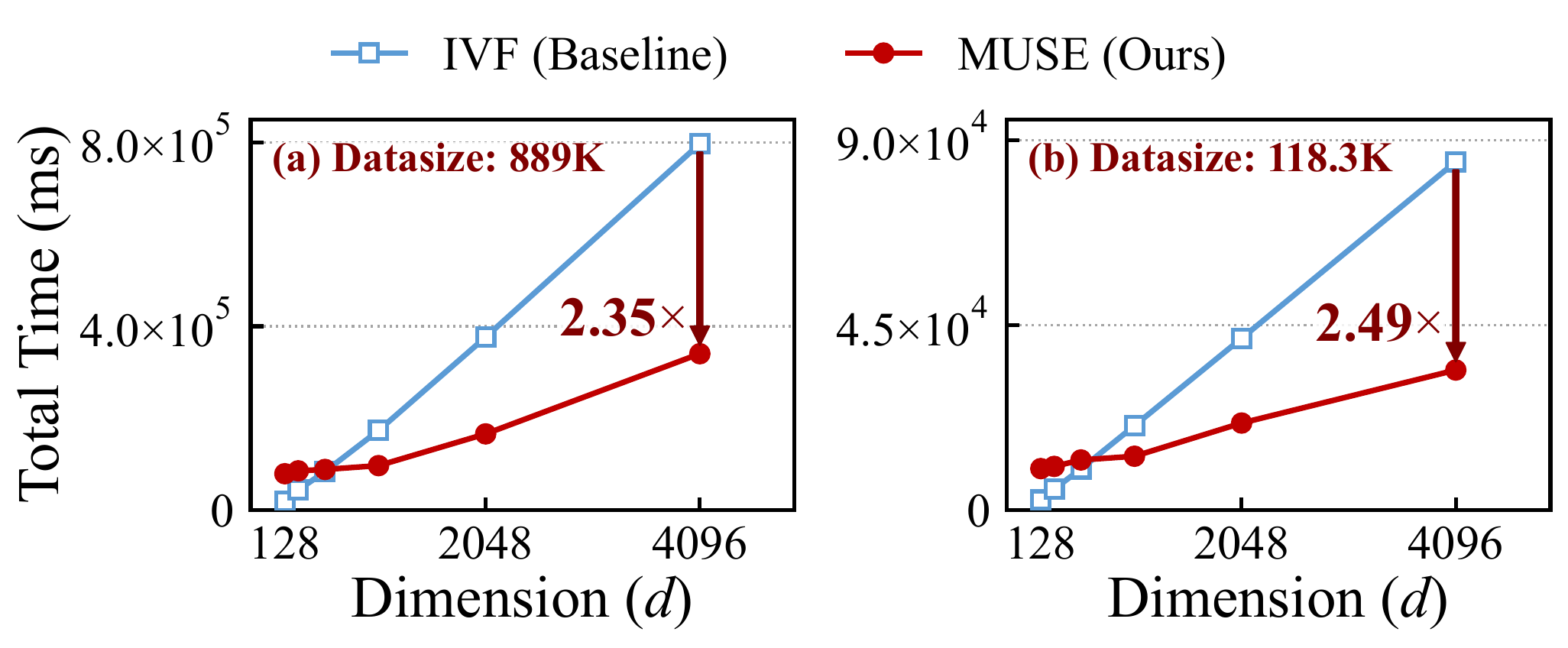}
    \caption{Scalability of total execution time across varying embedding dimensions.}
    \label{fig:multidim}
\end{figure}

As shown in Figure~\ref{fig:multidim}, the CPU-based IVF baseline suffers from a severe latency explosion at high dimensions due to unoptimized memory access patterns and limited vector execution width. In contrast, \ourmethod scales smoothly. By refactoring the search process into accelerator-native dense tile operations and overlapping DMA transfers, \ourmethod effectively mitigates the memory wall, achieving $2.35\times$ and $2.49\times$ speedups at $d=4096$ on datasets of 889K and 118.3K vectors, respectively. This demonstrates \ourmethod's readiness for next-generation, high-dimensional multimodal workloads.

\subsection{Energy and Thermal Efficiency}
On mobile SoCs, sustained high power draw triggers dynamic voltage and frequency scaling, causing severe throughput throttling~\cite{9044742}. To evaluate efficiency under these constraints, we profile total device power, mean temperature across multiple motherboard sensors, and energy footprint during index construction. All baselines are iso-configured for equivalent recall and query throughput.

\begin{figure}[t]
    \centering
    \includegraphics[width=1\linewidth]{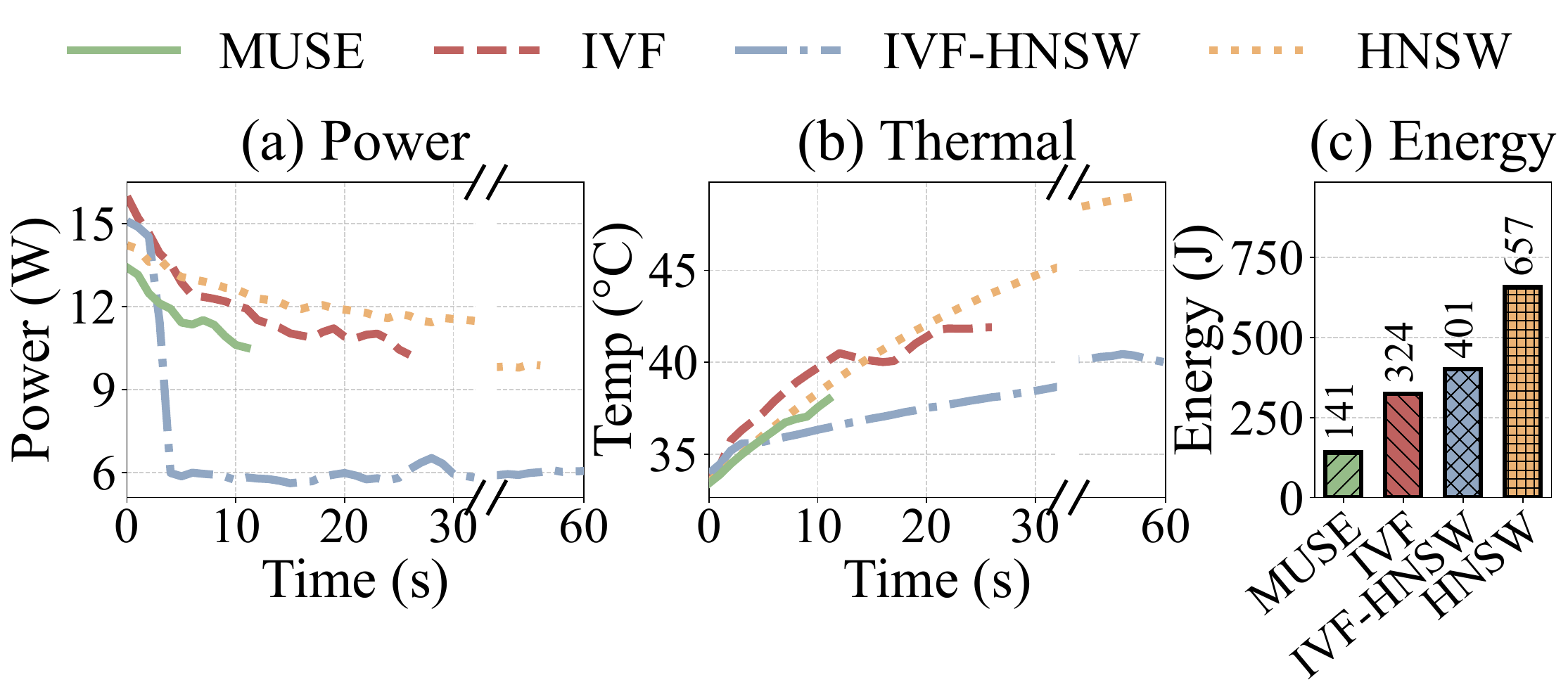}
    \caption{System-level efficiency profiles during index construction under mobile energy and thermal constraints: (a) power consumption, (b) thermal dynamics, and (c) total energy footprint.}
    \label{fig:energy}
\end{figure}

As shown in Figure~\ref{fig:energy}, \ourmethod completes the workload in 12 seconds. While engaging the concurrent heterogeneous pipeline induces a transient power peak ($\sim$13.4W), the execution concludes well before thermal saturation, capping device temperature at $38^\circ\text{C}$, well below the throttling threshold. Conversely, CPU-bound baselines prolong execution to 27--65 seconds; this sustained active load drives temperatures near $50^\circ\text{C}$, pushing the SoC into thermal throttling regimes.

By minimizing the active compute window, \ourmethod restricts total energy consumption to 141J—achieving $2.3\times$, $2.8\times$, and $4.6\times$ energy reductions over IVF (324J), IVF-HNSW (401J), and HNSW (657J), respectively. These results confirm that \ourmethod's accelerator-native design efficiently translates raw compute throughput into the strict thermal and energy compliance mandated by mobile platforms.

\section{Conclusion}
In this paper, we presented \ourmethod, a high-performance multimedia search engine comprehensively co-designed for the unique demands of mobile cross-modal applications. To resolve the performance paradox of naïve accelerator offloading for high-dimensional media embeddings, \ourmethod introduces an asynchronous execution pipeline that structurally overlaps zero-copy DMA transfers, on-the-fly layout adaptation, and NPU matrix computation. Coupled with a hardware-aligned index and a workload-aware heterogeneous scheduler, \ourmethod dynamically balances low-latency interactive multimodal queries with continuous, heavy background video/audio ingestion streams. Extensive evaluations on multiple generations of flagship mobile SoCs demonstrate that this system co-design substantially accelerates both massive on-device visual index construction and real-time retrieval. Furthermore, by minimizing the active compute window, \ourmethod{} translates raw execution speed into exceptional thermal and energy compliance, significantly reducing power consumption and preventing thermal throttling commonly observed in traditional CPU-bound approaches. Ultimately, \ourmethod proves that compute-intensive, high-dimensional multimedia lifelogging can operate efficiently on commodity, battery-constrained mobile endpoints, unlocking the privacy and responsiveness critical for next-generation multimodal AI agents.

\section{Limitations}
While \ourmethod's asynchronous pipeline and adaptation principles apply to modern NPUs (e.g., Hexagon~\cite{qualcomm_hexagon_arch, qualcomm_hexagon_hvx}, Apple ANE~\cite{apple2022ane}, MediaTek APU~\cite{mediatek2023apu}) due to their fundamentally similar architectural designs, our implementation targets Qualcomm hardware. Other vendors currently restrict NPU access to model-level APIs, precluding fine-grained DMA and ISA control required by our engine.

Furthermore, \ourmethod stores FP32 vectors in DRAM and dynamically downcasts them to FP16 via HVX, as our profiling reveals FP16 GEMM unexpectedly outperforms lower-precision variants on current silicon. Although FP32 increases memory footprint, our pipeline decouples storage from compute precision. This enables future integration of highly quantized vectors~\cite{gao2024rabitqquantizinghighdimensionalvectors} (e.g., FP8/INT4), asynchronously unpacking them to FP16 in TCM to reduce memory overhead without stalling execution.

\clearpage

\begin{acks}
This work was supported in part by the National Science Foundation of China under Grants (62472375), and in part by Zhejiang Provincial Natural Science Foundation of China under Grant No. LD24F020014 and No. LD25F020002, and in part by the Zhejiang Pioneer (Jianbing) Project (2024C01032), and in part by the Ningbo Yongjiang Talent Programme (2023A-198-G).
\end{acks}

\bibliographystyle{ACM-Reference-Format}
\bibliography{references}

\end{document}